\documentclass{article-hermes}

\usepackage{graphicx}

\usepackage{color}

\usepackage[bookmarks=true, bookmarksnumbered=true, bookmarksopen=true,
		unicode=true, colorlinks=true,
		linkcolor=blue,citecolor=blue,filecolor=blue,urlcolor=blue
		]{hyperref}

\begin{document}

\title[Rôle communautaire des capitalistes]%
      {Rôle communautaire des capitalistes sociaux dans Twitter}

\author{Nicolas Dugu\'e\fup{*} \andauthor Vincent Labatut\fup{**} \andauthor Anthony Perez\fup{*}}

\address{\fup{*}Université d'Orléans, LIFO\\
Bâtiment IIIA, Rue Léonard de Vinci, 45067 Orléans, France\\[3pt]
\{nicolas.dugue,anthony.perez\}@univ-orleans.fr\\[6pt]
\fup{**}Département d'informatique, Université Galatasaray\\
Ç{\i}ra\u{g}an cad. n°36, Ortaköy 34357, \.{I}stanbul, Turquie\\[3pt]
vlabatut@gsu.edu.tr}

\resume{Les capitalistes sociaux sont des utilisateurs de médias sociaux tels que Twitter, appliquant diverses techniques pour obtenir un maximum de visibilité. Ils peuvent \^etre n\'efastes à l'\'equilibre du service, dans la mesure o\`u leurs comptes, en gagnant en importance sans r\'eelle raison de contenu, rendent difficile l'accès à un contenu pertinent. Dans ce travail, nous nous intéressons à leur caractérisation d'un point de vue purement topologique, i.e. sans considérer la nature des contenus partagés. Nous utilisons pour cela la notion de rôle communautaire, qui est basée sur la structure de communautés du réseau étudié. Nous apportons des modifications à des mesures précédemment définies à cet effet, et proposons une méthode objective de détection des rôles. Nous appliquons ensuite notre méthode à l'analyse d'un réseau représentant Twitter. Nos résultats montrent que les r\^oles que nous identifions via nos mesures se r\'ev\`elent particuli\`erement coh\'erents par rapport aux capitalistes sociaux du r\'eseau Twitter, dont le comportement est clairement identifi\'e.}

\abstract{Social capitalists are social media users taking advantage of various methods to maximize their visibility. This results in artificially important accounts, in the sense this importance is not backed by any real content. The risk is then to see those accounts hiding relevant contents and therefore preventing other users to access them. In this work, we want to characterize social capitalists from a purely topological perspective, i.e. without considering the nature of the shared contents. For this purpose, we use the notion of community role, based on the community structure of the studied network. We modify some measures previously designed for this matter, and propose an objective method to determine roles. We then apply this method to the analysis of a Twitter network. Our results show the roles identified through our measures are particularly consistent with Twitter's social capitalists, whose behavior was clearly identified.}

\motscles{Réseau social, Twitter, Capitalisme social, Structure de communautés, Rôle.}


\proceedings{MARAMI 2013}{-}

\maketitlepage

\section{Introduction}
\noindent \textbf{Contexte.} Le principe du capitalisme social sur les r\'eseaux sociaux tels que Twitter consiste \`a essayer d'obtenir un maximum de visibilit\'e en utilisant diverses techniques. Cette notion a été mise en évidence dans un travail sur les comptes spammers de Twitter par Ghosh \textit{et al.}~\cite{GVK+12}. En ce qui concerne le r\'eseau social Twitter, les capitalistes sociaux utilisent deux principes relativement simples pour accroitre leur nombre de followers et ainsi leur visibilit\'e : 

\begin{itemize}
	\item \textbf{FMIFY} : Follow Me I Follow You, ils promettent aux comptes qui les suivent de les suivre en retour, 
	\item \textbf{IFYFM} : I Follow You, Follow Me, ils suivent un maximum de comptes, en esp\'erant \^etre suivi en retour.

\end{itemize}

De tels utilisateurs peuvent \^etre n\'efastes pour l'\'equilibre du r\'eseau social, dans la mesure o\`u leurs comptes gagnent en visibilit\'e et leurs tweets sont bien class\'es dans les moteurs de recherche du r\'eseau souvent sans r\'eelle raison de contenu. Notons \'egalement que de nombreux comptes populaires sur Twitter (tels que Barack Obama) sont connus pour avoir utilis\'e de tels principes. De ce fait, il est int\'eressant d'\^etre capable de d\'etecter ces utilisateurs de mani\`ere efficace. 

En se basant sur deux mesures simples, Dugu\'e \& Perez~\cite{DUGUE2013} ont r\'ecemment mis en avant que détecter ces capitalistes sociaux pouvait \^etre r\'ealis\'e en consid\'erant \emph{uniquement} le graphe des relations entre utilisateurs de Twitter. Dans un premier temps, il est utile de remarquer que les deux principes {\bf FMIFY} et {\bf IFYFM}, lorsqu'ils sont appliqu\'es aboutissent \`a un r\'esultat tr\'es similaire, \`a savoir qu'il y a une forte intersection entre les ensembles des followers et des followees. N\'eanmoins, le premier type de capitalistes devrait avoir un nombre de followers plus \'elev\'e que son nombre de followees, et inversement pour le second type. Ainsi, la notion d'\emph{indice de chevauchement}\footnote{De mani\`ere informelle, l'indice de chevauchement compare la taille de l'intersection des followers et des followees d'un utilisateur avec la taille du plus petit des deux ensembles.} a \'et\'e utilis\'ee pour d\'etecter les capitalistes sociaux sur le graphe des relations entre utilisateurs de Twitter, tandis que le \emph{ratio} (nombre de followees divis\'e par nombre de followers) permet de les caract\'eriser selon les deux types pr\'ec\'edemment cit\'es. 


Dans cet article, nous nous focalisons sur la notion de visibilité des n\oe{}uds d'un réseau social comme Twitter. Pour étudier cela, nous nous plaçons au niveau des communautés du réseau. La détection de communautés est une tâche qui, dans sa forme la plus simple, consiste à partitionner l'ensemble des n\oe{}uds constituant un réseau, de manière à faire apparaître des groupes de n\oe{}uds plus densément connectés entre eux qu'avec le reste du réseau \cite{Newman2004a}. Un très grand nombre de méthodes de détection de communautés a été développé récemment, cf. \cite{Fortunato2010} pour une revue détaillée du domaine. L'objectif de ce type d'analyse est de permettre d'étudier le réseau à un niveau \textit{intermédiaire}, comparé aux niveaux \textit{global} (réseau entier) ou \textit{local} (voisinage d'un n\oe{}ud) habituellement pris en compte. Considérons par exemple le cas d'une mesure caractérisant un n\oe{}ud pris individuellement. Alors, le degré $k$ du n\oe{}ud constitue une mesure locale, elle est basée sur la taille du voisinage direct. Au contraire, une mesure de centralité telle que l'intermédiarité (\textit{closeness} en anglais) correspond à une mesure globale, puisqu'elle est définie comme l'inverse de la distance géodésique moyenne entre le n\oe{}ud considéré et les autres n\oe{}ud du graphe \cite{Freeman1978}. Enfin, la mesure d'enchâssement $e$ (\textit{embeddedness} en anglais) proposée par Lancichinetti est située à un niveau intermédiaire, elle correspond à la proportion de liens qu'un n\oe{}ud $u$ possède avec les autres n\oe{}uds de sa propre communauté \cite{Lancichinetti2010}. Plus formellement, il s'agit du rapport entre le degré interne $k_{int}$ du n\oe{}ud (nombre de voisins appartenant à sa communauté) et son degré total $k$, i.e. $e(u)=k_{int}(u)/k(u)$.

Ainsi, pour caractériser la visibilité des n\oe{}uds dans le réseau, nous tentons d'établir leurs \emph{r\^oles} et \emph{positions} au sein de ces communautés. Notre objectif est de d\'eterminer si les n\oe{}uds sont ancr\'es au sein de leurs communaut\'es, \'etroitement li\'es avec les autres utilisateurs, ou s'ils sont au contraire isol\'es. De plus, nous consid\'erons \'egalement le comportement des n\oe{}uds par rapport aux autres communaut\'es, en nous demandant notamment s'ils sont li\'es aux autres communaut\'es, et, si oui, avec quelle intensit\'e. 
Cette analyse s'applique particulièrement aux capitalistes sociaux. En effet, ceux-ci sont susceptibles d'obtenir de la visibilité de façon artificielle.

\noindent \textbf{Contributions.} Du point de vue théorique, nous généralisons les travaux de Guimer\`a et Amaral~\cite{Guimera2005} portant sur la notion de \emph{r\^ole communautaire}, de deux manières distinctes. Premièrement, nous introduisons trois nouvelles mesures permettant de les caract\'eriser en fonction des connexions d'un n\oe{}ud au sein de sa communaut\'e et vers les communaut\'es ext\'erieures. En particulier, nos mesures tiennent compte de l'\emph{orientation} du graphe \'etudi\'e, ce qui se r\'ev\`ele pertinent du point de vue de l'analyse (Section~\ref{subsec:mesures}). Deuxièmement, nous effectuons une \emph{analyse de regroupement} sur ces diverses mesures, ce qui nous permet de g\'en\'erer des \emph{groupes} d'utilisateurs (Section~\ref{proprietesgroupes}), que nous interprétons comme des rôles. Cette approche est \emph{objective}, par opposition aux seuils originalement définis par Guimer\`a et Amaral de façon empirique. Du point de vue pratique, nous analysons un graphe mod\'elisant les relations entre utilisateurs de Twitter fourni par~\cite{CHBG10}. Nous identifions d'abord les capitalistes sociaux grâce à la méthode définie par Dugu\'e et Perez~\cite{DUGUE2013}. Il nous est alors possible d'étudier leur répartition au sein des groupes détectés, et de montrer qu'ils occupent certains rôles communautaires en particulier (Section~\ref{subsec:positionnement}).

\section{Rôle communautaire d'un n\oe{}ud}
Dans ce travail, nous nous intéressons à la caractérisation de la position des n\oe{}uds relativement à la structure de communautés. De ce point de vue, la mesure d'enchâssement de Lancichinetti est relativement limitée, dans le sens où elle ne permet de caractériser que grossièrement la position d'un n\oe{}ud dans sa propre communauté, et ignore complètement sa position vis-à-vis des autres communautés. L'approche proposée par Guimerà et Amaral semble plus adaptée à notre objectif \cite{Guimera2005}. Dans cette section, nous la décrivons et la critiquons dans un premier temps, puis nous proposons plusieurs modifications destinées à résoudre ses limitations.

\subsection{Degré intra-module et coefficient de participation}
Guimerà et Amaral proposent d'utiliser deux mesures traitant respectivement des aspects interne et externe à la communauté du n\oe{}ud. La première, nommée \textit{degré intra-module} (\textit{within-module degree} en anglais) est basée sur la notion de $z$-score. Comme celle-ci sera réutilisée plus loin, nous la définissons ici de façon générique. Pour une fonction nodale quelconque $f(u)$, permettant d'associer une valeur numérique à un n\oe{}ud $u$, le $z$-score $Z_f(u)$ par rapport à la communauté de $u$ est :
\begin{equation}
\label{f:zscore}
Z_f(u) = \frac{f(u) - \mu_i(f)} {\sigma_i(f)} 
\mbox{, avec } u \in C_i
\end{equation}
où $C_i$ représente une communauté, et $\mu_i(f)$ et $\sigma_i(f)$ dénotent respectivement la moyenne et l'écart-type de $f$ sur les n\oe{}uds appartenant à la communauté $C_i$. Le degré intra-module de Guimerà et Amaral, noté $z(u)$, correspond au $z$-score du degré interne, calculé pour la communauté du n\oe{}ud considéré. On l'obtient donc en substituant $k_{int}$ à $f$ dans l'équation [\ref{f:zscore}]. Cette mesure est sémantiquement très proche de l'enchâssement. En revanche, le degré intra-module évalue la connectivité d'un n\oe{}ud à sa communauté relativement à celle des autres n\oe{}uds de sa communauté. Cette notion n'apparaît pas dans l'enchâssement, qui relativise la connectivité interne par le degré total du n\oe{}ud. La seconde mesure, appelée \textit{coefficient de participation}, est définie de la manière suivante :
\begin{equation}
P(u) = 1 - \sum_i{\left(\frac{k_i(u)}{k(u)}\right)^2}
\end{equation}
où $k_c(u)$ représente le nombre de liens que $u$ possède vers des n\oe{}uds de la communauté $C_i$. Notons que dans le cas où $C_i$ est la communauté de $u$, alors on a $k_i(u) = k_{int}(u)$. Le coefficient de participation représente combien les connexions d'un n\oe{}ud sont diversifiées, en termes de communauté externes. Une valeur proche de $1$ signifie que le n\oe{}ud est connecté de façon uniforme à un grand nombre de communautés différentes. Au contraire, une valeur de $0$ ne peut être atteinte que si le n\oe{}ud n'est connecté qu'à une seule communauté (vraisemblablement la sienne).

Guimerà et Amaral~\cite{Guimera2005} proposent de caractériser le rôle d'un n\oe{}ud dans un réseau en se basant sur ces deux mesures. Pour ce faire, ils définissent sept rôles différents en discrétisant l'espace à deux dimensions formé par $z$ et $P$. Un premier seuil défini sur le degré intra-module $z$ permet de distinguer ce que les auteurs appellent les \textit{pivots communautaires} ($z\geq2.5$) des autres n\oe{}uds ($z<2.5$). Ces pivots (\textit{hubs} en anglais) sont considérés comme fortement intégrés à leur communauté, par rapport au reste des n\oe{}uds de cette même communauté. Ces deux catégories (pivot et non-pivot) sont subdivisées au moyen d'une série de seuils définis sur le coefficient de participation $P$. En considérant les n\oe{}uds par participation croissante, Guimerà et Amaral les qualifient de provinciaux ou (ultra-)périphériques, connecteurs et orphelins. Les deux premiers rôles sont essentiellement connectés à leur communauté, les troisièmes, bien qu'eux aussi potentiellement bien connectés à leur propre communauté, sont également largement liés à d'autres communautés, et les derniers sont seulement marginalement attachés à leur communauté.

\subsection{Approche proposée}
\label{subsec:mesures}
La notion de rôle définie dans \cite{Guimera2005} a été utilisée avec succès pour montrer que certains systèmes complexes possèdent des propriétés d'invariance de rôle : lorsqu'on considère plusieurs instances de système, les n\oe{}uds diffèrent mais les rôles restent distribués de façon similaire. Dans le cadre de notre travail, nous nous intéressons aux rôles occupés par les capitalistes sociaux. À cette fin, nous proposons plusieurs modifications de la méthode proposée par Guimerà et Amaral.

Considérons tout d'abord les mesures. Le coefficient de participation se concentre sur un aspect de la connectivité externe d'un n\oe{}ud : l'\textit{hétérogénéité} de la distribution de ses liens, relativement aux communautés auxquelles il est connecté. Mais il est possible de caractériser cette connectivité de deux autres manières. Premièrement, on peut considérer sa \textit{diversité}, c'est à dire le nombre de communautés concernées. Deuxièmement, il est possible de s'intéresser à son \textit{intensité}, i.e. au nombre de liens concernés. Ces deux aspects ne sont pas pris en compte dans $P$. Pour palier cette limitation, nous proposons deux nouvelles mesures permettant de quantifier la diversité et l'intensité. De plus, afin d'obtenir un ensemble cohérent de mesures, nous révisons également $P$.

\noindent \textbf{Diversité.} Notre mesure de \textit{diversité}, notée $D(u)$, \'evalue le nombre de communaut\'es diff\'erentes auxquelles le n\oe{}ud $u$ est connect\'e. 
, indépendamment de la densité de ces connexions. Soit $\epsilon(u)$ le nombre de communautés, autres que la sienne, auxquelles le n\oe{}ud $u$ est connecté. Alors la diversité est définie comme le $z$-score d'$\epsilon$ relativement à la communauté de $u$. C'est à dire qu'on l'obtient en substituant $\epsilon$ à $f$ dans [\ref{f:zscore}].

\noindent \textbf{Intensité externe.} L'\textit{intensité externe} $I_{ext}(u)$ mesure la force de la connexion de $u$ à des communautés externes, en termes de nombre de liens, et relativement aux autres n\oe{}uds de sa communauté. Soit $k_{ext}(u)$ le degré externe de $u$, correspondant au nombre de liens que $u$ possède avec des n\oe{}uds n'appartenant pas à sa communauté. Remarquons qu'on a alors $k=k_{int}+k_{ext}$. Nous définissons l'intensité externe comme le $z$-score du degré externe, c'est à dire qu'on l'obtient en substituant $k_{ext}$ à $f$ dans [\ref{f:zscore}].

\noindent \textbf{Hétérogénéité.} L'\textit{hétérogénéité} $H(u)$ quantifie combien le nombre de connexions externes du n\oe{}ud $u$ varie d'une communauté à l'autre. Nous utilisons pour cela l'écart-type du nombre de liens externes que le n\oe{}ud possède par communauté, que nous notons $\lambda(u)$. L'hétérogénéité est alors le $z$-score de $\lambda$, relativement à la communauté de \textit{u}, et on l'obtient donc en substituant $\lambda$ à $f$ dans [\ref{f:zscore}]. Cette mesure a une signification très proche de celle du coefficient de participation $P$ de Guimerà et Amaral. Elle diffère en ce qu'elle est exprimée relativement à la communauté de $u$, et que les liens internes à cette même communauté sont exclus du calcul.

\noindent \textbf{Intensité interne.} Pour représenter la connectivité interne du n\oe{}ud, nous conservons la mesure $z$ de Guimerà et Amaral. En effet, celle-ci est construite sur la base du $z$-score, et est donc cohérente avec les autres mesures définies pour décrire la connectivité externe. De plus, il n'est pas nécessaire de lui adjoindre d'autres, mesures, car les notions d'hétérogénéité et de diversité n'ont pas de sens ici (puisqu'on considère seulement une seule communauté). Cependant, en raison de sa symétrie avec notre intensité externe, nous désignons $z$ sous le nom d'\textit{intensité interne}, et la notons $I_{int}(u)$. 

Notre dernière contribution, en ce qui concerne les mesures, est de faire la distinction entre les liens qui sortent de la communauté et ceux qui y entrent. En effet, le réseau étudié étant orienté, il nous parait nécessaire d'exploiter au maximum l'information disponible. Pour chacune des 4 mesures présentées, nous utilisons donc en réalité deux variantes, chacun concernée uniquement par un seul type de lien. Nous obtenons finalement un total de 8 mesures différentes pour décrire la position de chaque n\oe{}ud par rapport à la structure de communautés.

Nous apportons également une autre modification à la méthode de Guimerà et Amaral, cette fois au niveau de la définition des rôles. Dans \cite{Guimera2005}, les seuils sur lesquels les rôles sont basés sont sélectionnés de façon arbitraire. Outre le fait que l'on peut critiquer cette approche en termes d'objectivité, il nous parait difficile de l'appliquer dans notre cas en raison du nombre élevé de mesures, qui complique l'estimation intuitive de ces seuils. Nous proposons plutôt d'effectuer une partition automatique de l'espace des mesures en réalisant une analyse de regroupement (\textit{cluster analysis} en anglais), une méthode issue de la fouille de données. Ainsi, le nombre des rôles et leur nature sont déterminés de façon objective.

\section{Résultats}
Le réseau sur lequel nous avons travaillé a été collecté en 2009 par \cite{CHBG10} et est accessible à tous. Il comporte un peu moins de $55$ millions de n\oe{}uds représentant les utilisateurs de Twitter et près de $2$ milliards d'arcs orientés qui matérialisent les abonnements entre utilisateurs, à savoir les liens de "follow". La très grande taille de ces données a influencé le choix de nos outils d'analyse. La détection de communautés a été réalisée au moyen de l'algorithme de Louvain \cite{Blondel2008}, car il s'est révélé très efficace dans le traitement de grands réseaux. Nous avons repris le code mis à disposition par ses auteurs et l'avons adapté à la modularité orientée décrite par Leicht \& Newman dans \cite{Newman2008}. Nos $8$ mesures ont également été implémentées en C++ avec, pour stocker le graphe, la même structure de matrice creuse que celle utilisée dans l'implémentation de Louvain. Elles ont été calculées sur la base des communautés identifiées par Louvain en utilisant la modularité orientée. Nous avons ensuite centré et réduit les données obtenues, afin d'éviter des problèmes de différence d'échelle lors de l'analyse de regroupement. Cette dernière a alors été menée au moyen d'une implémentation libre et distribuée de l'algorithme des $k$-moyennes \cite{Liao2009}. En effet, les méthodes non-distribuées, basées sur le calcul d'une unique matrice de distance, se sont révélées impossible à appliquer en raison de la quantité de mémoire nécessaire à la représentation de la matrice. Nous avons appliqué cet algorithme pour des valeurs de $k$ allant de $2$ à $15$, et avons sélectionné la meilleure partition d'après l'indice de Davies-Bouldin \cite{Davies1979}. Les scripts de pré- et post-traitement relatifs à l'analyse de regroupement on été implémentés en langage R. L'ensemble de notre code source est disponible à l'adresse 
\url{https://github.com/CompNet/Orleans} 
.

\subsection{Propriétés des groupes détectés}
\label{proprietesgroupes}
Considérons tout d'abord les mesures obtenues sur l'ensemble des données traitées. On observe des corrélations positives pour l'ensemble des paires de mesures, allant de valeurs proches de $0$ à $0,9$. Les deux variantes d'une même mesure (liens entrants contre liens sortants) sont peu corrélées, ce qui peut être expliqué par le découplage observé entre les degrés entrant et sortant. Trois mesures sont fortement corrélées : les intensités internes et externes et l'hétérogénéité ($\rho$ allant de $0,78$ à $0,92$). Le lien entre les intensités interne et externe semble indiquer que les variations dans le degré total d'un n\oe{}ud ont globalement le même effet sur ses degrés internes et externes. Autrement dit, la proportion entre ces deux types de liens ne dépend pas du degré du n\oe{}ud. Le très fort lien observé entre hétérogénéité et intensité indique que seuls les n\oe{}uds de faible intensité sont connectés de façon homogène à des communautés externes, tandis que les n\oe{}uds possédant de nombreux liens sont connectés de façon hétérogène.

En ce qui concerne l'analyse de regroupement, nous obtenons la meilleure séparation pour $k=6$ groupes, dont le Tableau \ref{tab:groupes} donne les tailles. Pour les interpréter, nous sommes partis de l'hypothèse que chaque groupe correspond à un rôle communautaire spécifique. Nous avons alors caractérisé les groupes relativement à nos huit mesures, afin d'en identifier les rôles et de les comparer à ceux définis par Guimerà et Amaral. Le Tableau \ref{tab:moyennes} contient les valeurs moyennes obtenues pour chaque mesure dans chaque groupe. Les {\sc ANOVA} que nous avons réalisées ont révélé des différences significatives pour toutes les mesures ($p<0.01$). Un test post-hoc ($t$-test avec correction de Bonferroni) a montré que ces différences existaient entre tous les groupes, pour toutes les mesures.

\begin{table}[h]
	\centering
	\begin{tabular}{|l|r|r|r|}
		\hline
		\textbf{Groupe} & \textbf{Taille} & \textbf{Proportion} & \textbf{Rôle} \\
		\hline
		1 & $24543667$ & $46,68\%$ & Non-pivot ultra-périphérique \\
		2 &      $304$ & $<0,01\%$ & Pivot orphelin \\
		3 &   $303674$ &  $0,58\%$ & Pivot connecteur	\\
		4 & $11929722$ & $22,69\%$ & Non-pivot périphérique (entrant) \\
		5 & $10828599$ & $20,59\%$ & Non-pivot périphérique (sortant) \\
		6 &  $4973717$ &  $9,46\%$ & Non-pivot connecteur \\
		\hline
	\end{tabular}
	\caption{Tailles de groupes détectés, et rôles correspondants dans la typologie de Guimerà et Amaral.}
	\label{tab:groupes}
\end{table}

Dans le groupe $1$, toutes les mesures sont négatives mais proches de $0$, à l'exception des deux variantes de la diversité, en particulier l'entrante, qui est proche de $-1$. Il ne peut pas s'agir de pivot au sens de Guimerà et Amaral (n\oe{}ud largement connecté  à sa communauté), puisque l'intensité interne est négative. De même, le fait que les mesures externes sont très faibles montre qu'il ne s'agit pas non plus de n\oe{}ud qualifiés de connecteurs par Guimerà et Amaral (ayant une connexion privilégiée avec d'autres communautés que la leur). On peut donc considérer que ce groupe correspond au role des non-pivots ultra-périphériques. Ce groupe est le plus grand (il contient à lui seul $47\%$ des n\oe{}uds), ce qui confirme la correspondance avec ce rôle, dont les n\oe{}uds constituent généralement la masse du réseau. Relativement au système modélisé, ces n\oe{}uds sont caractérisés par le fait qu'ils sont particulièrement peu suivis par les autres communautés.

Le groupe $4$ est extrêmement similaire au groupe $1$, à la différence que sa diversité entrante est de $0,69$. Ces n\oe{}uds restent donc périphériques, car l'intensité externe est toujours négative, mais ils reçoivent néanmoins des liens provenant d'un nombre relativement élevé de communautés. Autrement dit, ils sont suivis par peu d'utilisateurs externes, mais ceux-ci sont situés dans un grand nombres de communautés distinctes. Le groupe $5$ est lui aussi très proche du groupe $1$, mais la différence est cette fois que les deux variantes de la diversité sont positives, avec une diversité sortante de $0,60$. À l'inverse du groupe $4$, on peut donc dire ici que les utilisateurs concernés suivent (avec une faible intensité) des utilisateurs situés dans un grand nombre de communautés différentes. Les groupes $4$ et $5$ sont respectivement le deuxième ($23\%$) et troisième ($21\%$) plus grands groupes en termes de taille, ce qui porte le total des n\oe{}uds périphériques à $91\%$.

\begin{table}[h]
	\centering
	\begin{tabular}{|l|r|r|r|r|r|r|r|r|}
		\hline
		\textbf{G} &
 		\multicolumn{2}{|c|}{$\mathbf{I_{int}}$} & 
 		\multicolumn{2}{|c|}{$\mathbf{D}$} &
 		\multicolumn{2}{|c|}{$\mathbf{I_{ext}}$} &
 		\multicolumn{2}{|c|}{$\mathbf{H}$} \\
		\hline
		1 & $-0,12$ &  $-0,03$ & $-0,55$ & $-0,80$ &  $-0,09$ &  $-0,04$ &  $-0,12$ &  $-0,06$	\\
		2 & $94,22$ & $311,27$ &  $7,18$ & $88,40$ & $113,87$ & $283,79$ & $112,79$ & $285,57$	\\
		3 &  $5,52$ &   $1,40$ &  $5,60$ &  $3,10$ &   $5,28$ &   $1,43$ &   $6,76$ &   $2,34$	\\
		4 & $-0,04$ &   $0,00$ & $-0,37$ &  $0,69$ &  $-0,07$ &   $0,00$ &  $-0,10$ &  $-0,01$	\\
		5 & $-0,03$ &  $-0,01$ &  $0,60$ &  $0,19$ &  $-0,03$ &  $-0,02$ &  $-0,04$ &  $-0,02$	\\
		6 &  $0,48$ &   $0,12$ &  $1,96$ &  $1,70$ &   $0,35$ &   $0,12$ &   $0,53$ &   $0,19$	\\
		\hline
	\end{tabular}
	\caption{Mesures moyennes obtenues pour les $6$ groupes. Pour chaque mesure, deux valeurs sont indiquées, correspondant respectivement aux deux variantes : liens sortants et entrants.}
	\label{tab:moyennes}
\end{table}

Pour le groupe $6$, toutes les mesures sont positives. L'intensité interne reste proche de $0$, donc on ne peut toujours pas parler de pivot, même si ces n\oe{}uds sont mieux connectés à leur communautés que ceux des groupes précédents. L'intensité externe est elle aussi faible, mais le fait qu'elle soit positive, à l'instar des autres mesures externes, semble suffisante pour considérer ces n\oe{}uds comme des connecteurs au sens de Guimerà et Amaral (relativement bien reliés à d'autres communautés). La diversité est relativement élevée, aussi bien pour les liens entrants que sortants ($D>1,7$). Ces n\oe{}uds sont donc plus fortement connectés à leur communauté mais aussi à l'extérieur, et avec une plus grande diversité. Il s'agit du quatrième plus gros groupe, représentant $9,5\%$ des n\oe{}uds.

Toutes les mesures du groupe $3$ sont largement positives : supérieures à $1,4$ pour celles basées sur les liens entrants, et supérieures à $5,2$ pour les liens sortants. L'intensité interne élevée permet d'associer ce groupe au rôle de pivot. Les valeurs externes montrent en plus que ces n\oe{}uds sont connectés à de nombreux n\oe{}uds présents dans de nombreuses autres communautés. Toutefois, les liens sortants sont plus nombreux, ces n\oe{}uds correspondent donc à des utilisateurs plus suiveurs que suivis. Ce groupe ne représente que $0,6\%$ des n\oe{}uds, il s'agit donc d'un rôle bien plus rare que ceux associés aux groupes précédents. Cette observation est encore plus caractéristique du groupe $2$, qui représente bien moins de $1\%$ des n\oe{}uds. Toutes les mesures y sont particulièrement élevées, la plupart dépassant $100$. Pour une mesure donnée, la variante concernant les liens entrants est toujours largement supérieure, ce qui signifie que les utilisateurs représentés par ces n\oe{}uds sont particulièrement suivis, et donc influents. Nous associons ce groupe au rôle de pivot orphelin defini par Guimerà et Amaral.

En conclusion de cette analyse des groupes, on peut constater que tous les rôles identifiés par Guimerà et Amaral ne sont pas présents dans le réseau étudié : on n'y trouve ni non-pivots orphelins, ni pivots provinciaux. Cette observation semble confirmer la nécessité d'une approche objective pour déterminer comment regrouper les n\oe{}uds en fonction des mesures. Elle est également consistante avec la forte corrélation observée entre les intensités interne et externe : les rôles manquants correspondraient à des n\oe{}uds possédant une forte intensité interne mais une faible intensité externe, ou vice-versa. Or, ceux-ci sont très peu fréquents dans notre réseau. De plus, le fait de distinguer les liens entrants et sortants permet d'obtenir une typologie plus fine. Ainsi, certains groupes distincts ont émergé (groupes $4$ et $5$) là où l'approche de Guimerà et Amaral aurait considéré ces n\oe{}uds comme équivalents.

\subsection{Positionnement des capitalistes sociaux}
\label{subsec:positionnement}
Avec la m\'ethode définie dans \cite{DUGUE2013} et décrite en introduction, nous d\'etectons plus de $200000$ capitalistes sociaux. Nous \'etudions ici leur positionnement dans les $6$ groupes identifi\'es par la m\'ethode des $k$-moyennes. De plus, nous affinons notre analyse en structurant les capitalistes sociaux en diff\'erents groupes. Tout d'abord via le ratio, qui nous permet de mettre en évidence les comportements {\sc FMIFY} et {\sc IFYFM}. Ensuite, en utilisant le degr\'e de ces utilisateurs. En effet, les capitalistes sociaux ayant accru le plus efficacement leur nombre de followers sont susceptibles d'avoir un placement ou un r\^ole diff\'erent au sein des communaut\'es.

\noindent \textbf{Capitalistes sociaux de faible degré entrant.}

On s'intéresse aux capitalistes sociaux ayant un degré entrant compris entre $500$ et $10000$.
On découpe ce groupe en deux sous-groupes via le \emph{ratio} (le degré sortant du n\oe{}ud divisé par son degré entrant). On distingue ainsi les capitalistes sociaux ayant un ratio inférieur à $1$ (comportement de type FMIFY) et ceux de ratio supérieur à $1$ (comportement de type FMIFY).

Le tableau \ref{tab:ksociaux500} présente ainsi sur la première ligne la proportion de capitalistes sociaux du réseau qui sont contenus dans chaque groupe, et sur la deuxième la proportion de n\oe{}uds du groupe qui sont des capitalistes sociaux.
\begin{table}[h]
	\centering
	\begin{tabular}{|l|r|r|r|r|r|r|}
		\hline
		 \textbf{Ratio} & \textbf{G1}  & \textbf{G2} & \textbf{G3} & \textbf{G4} & \textbf{G5} & \textbf{G6}\\
		\hline
		  $<1$ & $0,03\%$ & $0,00\%$ & $\mathbf{14,64\%}$  &  $11,53\%$ &  $\mathbf{13,65\%}$ & $\mathbf{60,15\%}$  \\
		   & $< 0,01\%$ & $0,00\%$ & $\mathbf{4,29\%}$ &  $0,09\%$ &  $0,11\%$ & $1,07\%$  \\
		  \hline
		  $>1$  & $0,03\%$ & $0,00\%$ & $\mathbf{19,38\%}$  &  $0,48\%$ &  $\mathbf{14,07\%}$ & $\mathbf{66,05\%}$ \\
		    & $< 0,01\%$ & $0,00\%$ & $\mathbf{7,31\%}$ &  $< 0,01\%$ &  $0,14\%$ & $1,52\%$ \\
		\hline
	\end{tabular}
	\caption{Répartition des capitalistes sociaux de faible degré dans les différents groupes.}
	\label{tab:ksociaux500}
\end{table}

Ces capitalistes sociaux se regroupent dans trois groupes : $3$, $5$ et $6$. Les n\oe{}uds du groupe $3$ sont des pivots connecteurs, qui ont en particulier tendance à suivre plus d'utilisateurs du réseau que la normale. Ainsi, il semble naturel d'y voir apparaître un grand nombre de capitalistes sociaux, et particulièrement ceux de type {\sc IFYFM}, de ratio supérieur à $1$, qui suivent de nombreux autres utilisateurs dans l'espoir d'\^ etre suivis en retour. La diversité sortante élevée du groupe $3$ nous apprend également que ces capitalistes sociaux ont tendance à ne pas cibler uniquement leur communauté m\^eme s'ils y sont bien connectés, mais à appliquer leurs méthodes à travers de nombreuses communautés du réseau.

On observe que la large majorité des capitalistes sociaux de faible degré se place au sein du groupe $6$, non-pivot connecteur. Ces n\oe{}uds, qui sont légèrement plus connectés au sein de leur communauté et avec l'extérieur que la moyenne, ont en revanche une diversité bien plus élevée. Les capitalistes sociaux qui s'y situent semblent ainsi avoir débuté l'application de leurs méthodes, en créant des liens avec de nombreuses autres communautés.

Enfin, on retrouve une faible proportion de capitalistes sociaux de faible degré dans le groupe $5$, groupe de n\oe{}uds non-pivots périphériques. Un certain nombre de capitalistes sociaux sont ainsi isolés au sein de leur communauté et avec l'extérieur.

\noindent \textbf{Capitalistes sociaux de degré entrant élevé.}

On s'intéresse maintenant aux capitalistes sociaux ayant un degré supérieur à 10000.
On découpe ce groupe en trois sous-groupes via le \emph{ratio}. On distingue ainsi les capitalistes sociaux ayant un ratio inf\'erieur \`a $0,7$ que \cite{DUGUE2013} d\'ecrivent comme passifs, ayant arr\^et\'e d'appliquer les méthode de capitalisme social, ceux de ratio compris entre $0,7$ et $1$ dits FMIFY, et ceux de ratio sup\'erieur à $1$, de type IFYFM.

Le tableau suivant présente comme précédemment, sur la première ligne la proportion de capitalistes sociaux du réseau qui sont contenus dans chaque groupe, et sur la deuxième ligne, la proportion de n\oe{}uds du groupe qui sont des capitalistes sociaux.
\begin{table}[h]
	\centering
	\begin{tabular}{|l|r|r|r|r|r|r|}
		\hline
		\textbf{Ratio} & \textbf{G1}  & \textbf{G2} & \textbf{G3} & \textbf{G4} & \textbf{G5} & \textbf{G6}\\
		\hline
		   $<0,7$ & $0,00\%$ & $10,43\%$ & $\mathbf{81,67\%}$ & $0,00\%$ & $0,00\%$ & $7,90\%$ \\
		   & $0,00\%$ & $\mathbf{31,25\%}$ & $0,24\%$ &  $0,00\%$ &  $0,00\%$ & $< 0,01\%$ \\
		   \hline
		    $>0,7$ et $<1$ & $0,00\%$ & $1,52\%$ & $\mathbf{95,72\%}$ & $0,00\%$ & $0,00\%$ & $2,76\%$ \\
		    & $0,00\%$ & $7,24\%$ & $0,46\%$ &  $0,00\%$ &  $0,00\%$ & $< 0,01\%$ \\
		   \hline
		    $>1$ & $0,00\%$ & $0,03\%$ & $\mathbf{98,02\%}$ & $0,00\%$ & $0,00\%$ & $1,96\%$ \\
		    & $0,00\%$ & $0,33\%$ & $1,24\%$ &  $0,00\%$ &  $0,00\%$ & $< 0,01\%$ \\
		
		\hline
	\end{tabular}
	\caption{Répartition des capitalistes sociaux de degré élevé dans les différents groupes.}
	\label{tab:ksociaux10000}
\end{table}
Les capitalistes sociaux de degré élevé se retrouvent dans les groupes $2$ et $3$. Cela semble normal, ces groupes contiennent des n\oe{}uds pivots connecteurs et orphelins.

Comme détaillé précédemment, les n\oe{}uds du groupe $3$ sont des pivots connecteurs, qui suivent plus d'utilisateurs du réseau que la normale. Ainsi, on y observe à nouveau un grand nombre de capitalistes sociaux de degré élevé, et particulièrement ceux de type {\sc IFYFM}, de ratio supérieur à $1$, qui suivent de nombreux autres utilisateurs dans l'espoir d'\^ etre suivis en retour.

Les valeurs des mesures tenant compte des liens entrants des n\oe{}uds du groupe $2$ sont très élevées (Section \ref{proprietesgroupes}). Ces utilisateurs sont donc très suivis par un grand nombre de communautés différentes, de façon très intense, mais avec une distribution des liens hétérogène. On observe très nettement la présence d'un grand nombre de capitalistes sociaux de degré élevé de ratio inférieur à $0,7$. Ceux-ci représentent près de $31,25\%$ du groupe. Ce sont ceux qui ont le mieux réussi à acquérir de la visibilité sur le réseau. Il semble donc naturel de les retrouver au sein de ce groupe.


\section{Conclusion}
Dans cet article, notre but est d'étudier le rôle communautaire des capitalistes sociaux dans Twitter. Nous avons d'abord appliqué une méthode conçue précédemment pour identifier les capitalistes sociaux \cite{DUGUE2013}. Nous avons ensuite proposé une extension de la méthode définie dans \cite{Guimera2005} pour identifier les rôles communautaires dans un réseau complexe. Nos modifications avaient les objectifs suivants : 1) considérer les différents aspects de la connectivité d'un n\oe{}ud à des communautés externes (diversité, intensité et hétérogénéité), 2) distinguer les liens entrants et sortants et 3) identifier les rôles de façon objective. Notre méthode met en lumière les r\^oles caractéristiques joués par les capitalistes sociaux. La prise en compte de l'orientation des liens, notamment, permet d'obtenir des r\^oles plus pertinents, ce qui confirme l'intérêt d'exploiter cette information lors de l'\'etude des r\'eseaux sociaux.

Le travail présenté peut s'étendre de différentes façons. Tout d'abord, certains des rôles définis dans \cite{Guimera2005} n'apparaissent pas dans notre analyse. Il serait intéressant d'étudier d'autres réseaux afin de déterminer si cette observation reste valable. 
Une autre piste consiste à baser nos calculs sur des communautés recouvrantes (i.e. non-mutuellement exclusives). En effet, les réseaux sociaux que nous étudions sont réputés posséder ce type de structures, dans lesquelles un n\oe{}ud peut appartenir à plusieurs communautés en même temps \cite{Arora2012} ; de plus, de nombreux algorithmes existent pour les détecter \cite{Xie2013}. L'adaptation de nos mesures à ce contexte se ferait naturellement, en définissant des versions internes de l'hétérogénéité et de la diversité.

\bibliography{marami13}

\end{document}